\documentclass[twocolumn,prd,showpacs,nofootinbib]{revtex4}

\usepackage{graphicx}
\usepackage{textcomp}
\usepackage[usenames,dvipsnames]{color}
\usepackage{amsmath}
\usepackage{slashed}

\def\beq{\begin{equation}}
\def\eeq{\end{equation}}
\def\theta{\vartheta}

\newcommand{\be}{\begin{equation}}
\newcommand{\ee}{\end{equation}}
\newcommand{\ba}{\begin{eqnarray}}
\newcommand{\ea}{\end{eqnarray}}
\newcommand{\lsim}   {\mathrel{\mathop{\kern 0pt \rlap
  {\raise.2ex\hbox{$<$}}}
  \lower.9ex\hbox{\kern-.190em $\sim$}}}
\newcommand{\gsim}   {\mathrel{\mathop{\kern 0pt \rlap
  {\raise.2ex\hbox{$>$}}}
  \lower.9ex\hbox{\kern-.190em $\sim$}}}

\begin{document}


%
%







\title{Searching for Traces of Planck-Scale Physics with High Energy Neutrinos}
 
\author{Floyd W. Stecker}
\affiliation{Astrophysics Science Division, NASA Goddard Space Flight Center, Greenbelt, MD 20771, USA } 
\affiliation{Department of Physics and Astronomy, University of California at Los Angeles,
Los Angeles, CA 90095}
\author{Sean T. Scully}
\affiliation{Department of Physics and Astronomy, James Madison University \\ Harrisonburg, VA 22807, USA}

\author{Stefano Liberati}
\affiliation{SISSA - International School for Advanced Studies \\ Via Bonomea 265, Trieste 34136, Italy}
\affiliation{INFN Sezione di Trieste, Trieste, Italy}
\author{David Mattingly}
\affiliation{Department of Physics, University of New Hampshire \\ Durham, New Hampshire 03824, USA}

\begin{abstract}
High energy cosmic neutrino observations provide a sensitive test of Lorentz invariance violation (LIV), which may be a consequence of quantum gravity theories. We consider a class of non-renormalizable, Lorentz invariance violating operators that arise in an effective field theory (EFT) description of Lorentz invariance violation in the neutrino sector inspired by Planck-scale physics
and quantum gravity models. We assume a conservative generic scenario for the redshift distribution of extragalactic neutrino sources and employ Monte Carlo techniques to describe superluminal neutrino propagation, treating kinematically allowed energy losses of superluminal neutrinos caused by both vacuum pair emission (VPE) and neutrino splitting. We consider EFTs with both non-renormalizable $\cal{CPT}$-odd and non-renormalizable $\cal{CPT}$-even operator dominance. We then compare the spectra derived using our Monte Carlo calculations in both cases with the spectrum observed by IceCube in order to determine the implications of our results regarding Planck-scale physics. We find that if the drop off in the neutrino flux above $\sim 2$ PeV is caused by Planck scale physics, rather than by a limiting energy in the source emission, a potentially significant pileup effect would be produced just below the drop off energy in the case of $\cal{CPT}$-even operator dominance. However, such a clear drop off effect would {\it not} be observed if the  
$\cal{CPT}$-odd, $\cal{CPT}$-violating term dominates. 

\end{abstract}


\maketitle

\section{Introduction}

General relativity has been a fundamental tenet of physics for almost a century. Similarly, quantum field theory has also proved crucial to a deep understanding of physics, both as a fundamental framework to describe subatomic particles, and as a framework that describes emergent phenomena in condensed matter. However, merging the two theories naively yields an incomplete theory at the Planck scale of $\lambda_{Pl} = \sqrt{G\hbar/c^3} \sim 10^{-35}$ m~\cite{pl99} as general relativity is not perturbatively renormalizable. In their efforts to provide a UV (i.e., high energy) completion for quantum general relativity, many quantum gravity theories introduce drastic modifications to space-time at the Planck scale (e.g.,~\cite{do95}). Examples of this are extra dimensions theories and postulating fundamental discreteness of space-time, with the building blocks of nature being extended objects. 
 
One possible modification to space-time structure that has received quite a bit of attention is the idea that Lorentz symmetry is not an exact symmetry of nature.  Such a proposal is rather tame when compared with other quantum gravity ideas as historically the symmetry groups used to model physical phenomena have inevitably evolved over time. Lorentz symmetry violation has been explored within string theory~\cite{ko89}, loop quantum gravity, Ho\v{r}ava-Lifshitz gravity, causal dynamical triangulations, non-commutative geometry, doubly special relativity, among others (see, e.g., Refs.~\cite{dm05} and~\cite{li13} and references therein).

While it is not possible to directly investigate space-time physics at the Planck energy of $\sim 10^{19}$ \ GeV, many lower energy testable effects have been predicted to arise from the violation of Lorentz invariance (LIV) at or near the Planck scale. The subject of investigating LIV has therefore generated much interest in the particle physics and astrophysics communities. 

In this paper we propose using high energy astrophysical neutrino data to search for traces of such Planck-scale physics. We use the IceCube data~\cite{aa14} to investigate the possible effects of LIV terms arising within the context of an effective field theory (EFT) such as generalized in the standard model extension (SME) formalism. We concentrate on the lowest order Planck-mass suppressed operators, {\it viz.}, the mass dimension $[d] = 5$ and $[d] = 6$ terms that arise in an EFT description of LIV in the neutrino sector, showing the effect that these terms produce on the propagation of extragalactic high energy neutrinos. We then discuss the specific implications of our results, placing improved limits on the strength of the $[d] = 6$ operator and ruling out dominance of $\cal{CPT}$ violation from a $[d] = 5$ five operator. 

\section{Free particle propagation and modified kinematics}

In the effective field theory (EFT) formalism, LIV can be incorporated by the addition of
terms in the free particle Lagrangian that explicitly break Lorentz invariance. 
Since it is well known that Lorentz invariance holds quite well at accelerator
energies, the extra LIV terms in the Lagrangian must be very small. The EFT is considered
an approximation to a true theory that holds up to a limiting high energy (UV) scale.

\subsection{Mass dimension [d] = 4 LIV with rotational symmetry}

For an introduction as to how LIV terms affect particle kinematics,
we consider the simple example of a free scalar particle Lagrangian with an additional small
dimension-4 Lorentz violating term, assuming rotational symmetry~\cite{co99} (see also
Section IIB).
 
\begin{equation}
\Delta \mathcal{L}_f = \partial_{i}\Psi^{*}{\bf \epsilon}\partial^{i}\Psi.
\end{equation}
This leads to a modified propagator for a particle of mass $m$
\begin{equation}
-iD^{-1}~=~ (p_{(4)}^2~-~m^2)~+~\epsilon p^2.
\end{equation}
so that we obtain the dispersion relation
\begin{equation}
p_{(4)}^2~=~E^2~-~p^2~ \Rightarrow~ m^2 ~ +~ \epsilon p^2.
\label{LIVdispersion}
\end{equation}

In this example, the low energy "speed of light" maximum attainable particle velocity $(c = 1)$, here equal to 1 by convention, is replaced by a new maximum attainable velocity (MAV) as $v_{MAV} \neq 1$, which is 
changed by $\delta v \equiv \delta = \epsilon/2.$ 
\begin{equation}
{{\partial  E}\over{\partial |\vec{p}|}}  = {{|\vec{p}|}  \over {\sqrt
{|\vec{p}|^2 + m^2 v_{MAV} ^2}}} v_{MAV},
\label{groupvel}
\end{equation}
\noindent which goes  to $v_{MAV}$ at relativistic energies, $|\vec{p}|^2 \gg m^2$.

For the [d] = 4 case, the superluminal velocity of particle $I$ that is produced by the existence of one or more LIV terms in the free particle Lagrangian will be denoted by
\begin{equation}
v_{I, MAV} \equiv 1 + \delta_{I}
\label{v}
\end{equation}
We are always in the relativistic limit $|\vec{p}|^2 \gg m^2$ for both neutrinos and electrons. Thus, the neutrino or electron velocity is just given by equation~(\ref{v}). We note that in the case where $[d] > 4$ Planck-suppressed operators dominate, there will be LIV terms that are proportional to $(E/M_{Pl})^n$, where $n = [d] - 4$, leading to values of $\delta_{I}$ that are energy dependent (see next section).

\subsection{Fermion operators in standard model extension effective field theory}

Colladay and Kosteleck\'{y}~\cite{ck98} proposed a comprehensive EFT framework for quantifying and cataloging the empirical effects of small violations of $\cal{CPT}$ and Lorentz invariance known as the standard model extension (SME). The SME adds all possible Lorentz violating operators to the standard model that preserve the internal gauge symmetries and hence it is, in some sense, the most general possible model.  The total SME consists of hundreds of operators, many of which are very tightly constrained.  

The full SME can be simplified by imposing that subgroups of the Lorentz group or discrete symmetries such as $\cal{CPT}$ be preserved.  One common and useful simplification is that rotational invariance is still a good symmetry of nature in one particular frame (e.g.,~\cite{co99}). It is natural to take this frame to be the rest frame of the 2.7 K cosmic background radiation (CBR), a special frame picked out by the universe itself. Our motion with respect to this frame is only $\beta \simeq 10^{-3}$.  We will assume rotation invariance is preserved in the CBR frame for the rest of this paper and that our slight motion with respect to this frame will not significantly affect our results.

Since the rotation subgroup is compact it can be explored much more thoroughly than the boost subgroup, which is non-compact.  Hence rotation symmetry is much more tightly experimentally constrained. For example, Ho\v{r}ava-Lifshitz gravity~\cite{ho09}, one of the most popular quantum gravitational models that breaks Lorentz symmetry, postulates a preferred foliation for space-time but no other additional geometric structure.  In the reference frame associated with the foliation, rotation invariance is therefore preserved, as the directions associated with motion along a leaf of the foliation are indistinguishable. 

With the assumption of rotational invariance the number of possible operators for cosmic neutrinos drops significantly.  In addition, since we will be considering the effects of Lorentz violation on freely propagating cosmic neutrinos we only need to examine Lorentz violating modifications to the neutrino kinetic terms.  All such terms for Dirac fermions 
can be written by coupling derivatives of the fermion wave function to a unit norm vector field $u^a$, which defines the preferred frame. (Majorana couplings are ruled out in SME in the case of rotational symmetry~\cite{ko12}. Hence, for consistency we will also assume a [d] = 4 Dirac mass term for neutrinos.)  In natural units $\hbar=c=1$ the additional terms of interest up to mass dimension six that generate the corresponding lowest order corrections to the propagation of a free fermion $\psi$ are
\begin{eqnarray} 
\label{eq:Lagrangian}
\Delta \mathcal{L}_f= - M b\bar{\psi} \gamma_5(u \cdot \gamma) \psi \\ \nonumber
- i \bar{\psi}  (u\cdot \gamma) (d_L P_L + d_R P_R) (u \cdot D) \psi \\ \nonumber
+M^{-1} \bar{\psi}(e_L P_L + e_R P_R) (u \cdot \gamma) (u \cdot D)^2 \psi \\ \nonumber
 - M^{-1} \bar{\psi} (u \cdot D)^2 (f_L P_L + f_R P_R) \psi\\ \nonumber
 - i M^{-2} \bar{\psi} (u \cdot D)^3 (u \cdot \gamma) (g_L P_L + g_R P_R) \psi.
\end{eqnarray}
Here $P_{L,R}$ are the chiral projection operators $2 P_{L,R}=(1 \mp \gamma_5)$, $D$ is the gauge covariant derivative, and $M$ is a length scale, presumably set by quantum gravity. We will take $M$ to be the Planck energy $M_{Pl}$ for the rest of this paper. The corresponding dimensionless coefficients $b, d_{L,R},e_{L,R}, f_{L,R}, g_{L,R}$ can in principle be different for each fermion species and are what can be constrained by experiment.  

There are many ways the above LIV modifications to the Lagrangian can affect neutrino physics.  They give rise to an energy dependent effective neutrino mass, and so change the patterns of neutrino oscillations.  They also introduce corrections to the matrix elements for existing interactions as well as create new interactions between standard model fermions and $u^a$.  However, for our purposes, the most important effect these terms have is to change the kinematics of particle interactions with matrix elements governed by existing standard model physics. Since the Lorentz violating operators change the free field behavior and dispersion relation, interactions such as fermion-antifermion pair emission by neutrinos become kinematically allowed~\cite{co99} and can cause significant observational effects if the neutrinos are slightly superluminal. An example of such an interaction is electron neutrino splitting $\nu_e \rightarrow \nu_e + \nu_i + \bar{\nu_i}$ where $i$ is a flavor index.  Neutrino splitting can be represented as a rotation of the Feynman diagram for neutrino-neutrino scattering  which is allowed by relativity.  However, absent a violation of Lorentz invariance, neutrino splitting is forbidden by conservation of energy and momentum.  As we shall see, the dominant pair emission reactions are neutrino splitting and its close cousin, vacuum electron-positron pair emission (VPE) $\nu_i \rightarrow \nu_i + e^+ + e^-$, as these are the reactions with the lightest final state masses. We now set up a simplified formalism to calculate the observational effect of these two specific anomalous interactions on the neutrino spectrum seen in IceCube.

Varying the standard Dirac Lagrangian with the extra Lorentz violating terms in equation (\ref{eq:Lagrangian}) with respect to $\psi$ and looking for wave solutions with definite helicity $s=\pm1$ yields the corresponding species dependent particle dispersion relation.  At high energies, assuming that the Lorentz violating terms yield small corrections to $E$ and $p$, it follows that $E \simeq p$ and one can treat helicity and chirality as degenerate. We then find the dispersion relation
\begin{eqnarray} 
\label{eq:dispersion}
E^2-p^2=m^2 +  (1-s)\left(d_L p^2 + e_L\frac{p^3}{M_{Pl}} + g_L \frac{p^4}{M_{Pl}^2}\right) \\  \nonumber
+ (1+s)\left(d_R p^2 + e_R\frac{p^3}{M_{Pl}} + g_R \frac{p^4}{M_{Pl}^2}\right) \\ \nonumber + \frac{m}{M_{Pl}}(f_L + f_R) p^2 + f_L f_R \frac{p^4}{M_{Pl}}.
\end{eqnarray}
It follows from equation~(\ref{eq:dispersion}) that we are assuming a power expansion in momentum
with $M_{Pl}$ taken as the UV scale that fixes its domain of validity to be $p \le M_{Pl}$. As the above relation makes clear, multiple coefficients in the Lagrangian yield the same kind of dispersion modification, with the deviations scaling as $E^{2+n}$ for $n = 0,1,2$.  Free particle observations therefore cannot directly test a single Lorentz violating coefficient.  In addition, observable Lorentz violating effects from anomalous particle interactions generally depend on combinations of the coefficients for different species.  

A useful simplified formalism for analyzing such kinematics that highlights the physical process is to wrap the additional Lorentz violating terms into an effective mass term, $\tilde{m}_I(E)$, which is the right hand side of equation (\ref{eq:dispersion}) labeled by a particle species index $I$.  We can further identify $\tilde{m}(E)$ using equation (\ref{LIVdispersion}), yielding the relation
\begin{equation} 
\tilde{m}^2(E)=m^2+ 2\delta_I E^2,
\label{effectivemass}\end{equation}
where the velocity parameters $\delta_I$ are now energy dependent dimensionless ($c = 1$) coefficients for each species that can be directly identified from the fundamental parameters in the Lagrangian.  Therefore constraining $\delta_I$ for a particle provides limits on the fundamental Lorentz violating Lagrangian. Similarly, we define the parameter $\delta_{IJ} \equiv \delta_{I} - \delta_{J}$ as the Lorentz violating difference between the velocities of particles $I$ and $J$. In general $\delta_{IJ}$ will therefore be of the form
\begin{equation}
\delta_{IJ}=\sum_{n=0,1,2}\kappa_{IJ,n} \left(\frac{E}{M_{Pl}}\right)^n.
\label{d}
\end{equation}

The $\kappa_{\nu e,0}$ coefficient has already been tightly constrained~\cite{st14} from the observation of extraterrestrial PeV scale neutrinos by the IceCube collaboration~\cite{aa14}. If we wish to assume the dominance of Planck-suppressed terms in the Lagrangian as tracers of Planck scale physics, we make the assumption here that $\kappa_{\nu e,0} \ll \kappa_{\nu e,1}, \kappa_{\nu e,2}$\footnote{Several mechanisms have been proposed for the suppression of the LIV $[d] = 4$ term in the
Lagrangian. See, e.g., the review in Ref.~\cite{li13}.} 
Alternatively, we may postulate the existence of only Planck-suppressed terms in the Lagrangian, i.e., $\kappa_{\nu e,0} = 0$. We can further simplify by noting the important connection between LIV and $\cal{CPT}$ violation. Whereas a local interacting theory that violates $\cal{CPT}$ invariance will also violate Lorentz invariance~\cite{gr02}, the converse does not follow; an interacting theory that violates Lorentz invariance may, or may not, violate $\cal{CPT}$ invariance. LIV terms of odd mass dimension $[d]= 4 + n$ are $\cal{CPT}$-odd and violate $\cal{CPT}$, whereas terms of even mass dimension are $\cal{CPT}$-even and do not violate $\cal{CPT}$~\cite{ko09}. We can then specify a dominant term for $\delta_{IJ}$ in equation (\ref{d}) depending on our choice of $\cal{CPT}$. Considering Planck-mass suppression, the dominant term that admits $\cal{CPT}$ violation is the $n = 1$ term in equation (\ref{d}). On the other hand, if we require $\cal{CPT}$ conservation, the
$n = 2$ term in equation (\ref{d}) is the dominant term. Thus, we can choose as a good approximation to equation (\ref{d}), a single dominant term with one particular power of $n$ by specifying whether we are considering $\cal{CPT}$ even or odd LIV.  As a result, $\delta_{IJ}$ reduces to
\begin{equation}
\delta_{IJ}~ \equiv \kappa_{IJ,n} \left({{E}\over{{M_{Pl}}}}\right)^{n} 
\label{sub}
\end{equation}
with $n = 1$ or $n = 2$ depending on the status of $\cal{CPT}$.  Constraints are therefore most directly expressed in terms of limits on $\kappa_{\nu e,1}$ and $\kappa_{\nu e,2}$. For $[d] > 4$ the superluminal velocity excesses are given as integral multiples of $\kappa_{\nu e,1}$ and $\kappa_{\nu e,2}$ through the group velocity relation given by equation~(\ref{groupvel}). 
We note that in the SME formalism, since odd-[d] LIV operators are $\cal{CPT}$ odd, the $\cal{CPT}$-conjugation property implies that neutrinos can be superluminal while antineutrinos are subluminal or vice versa~\cite{ko12}. This will have consequences in interpreting our results, as we will discuss later.

We note that the $\nu$ is used here generically for all three neutrino flavors, $\nu_e, \nu_\mu$, and $\nu_{\tau}$  We have also put no helicity index on $\kappa_{IJ,n}$.  Since the fundamental parameters in the Lagrangian are helicity dependent we have made an additional, {\it a priori} unjustified simplification.  Let us first deal with the issue of helicity dependence in $\kappa_{\nu e, n}$.  For processes mediated by standard model matrix elements, only left-handed neutrinos can be constrained. Therefore we are insensitive to $d_R,e_R,f_R,g_R$ in the neutrino sector and can never generate a helicity dependence this way.  In the $n = 1$ case, a helicity dependence must be generated in the electron sector due to the $\cal{CPT}$ odd nature of the LIV term, but the constraints on the electron coefficient are extremely tight from observations of the Crab nebula~\cite{ja03,st14b},, and so the contribution to $\kappa_{\nu e,1}$ from the electron sector can be neglected.  Therefore there is no helicity dependence in $\kappa_{\nu e,1}$.  For $n = 2$ we can set the left and right handed electron coefficients to be equal by imposing parity symmetry, which we do here.

We will further assume that all neutrino flavors have the same LIV coefficient, $\delta_{\nu}$.
This assumption is supported by neutrino oscillation results that find that velocity
differences among neutrino flavors are equal to within one part in $10^{22}$~\cite{ab14}. 

\section{Limits on LIV in the neutrino sector}

\begin{figure}[htb]
\includegraphics[scale=0.35]{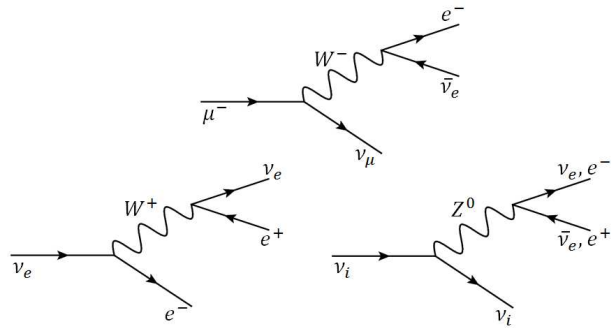}
\caption{Diagrams for muon decay (top), charged current mediated VPE (bottom left), and neutral current mediated neutrino splitting/VPE (bottom right).  Time runs from left to right and the flavor index $i$ represents $e,\mu$, or $\tau$ neutrinos.}\label{fig:diagrams}
\end{figure}

In this section we consider the constraints on the LIV parameter $\delta_{\nu e}$. We first relate the rates for superluminal neutrinos with that of a more familiar tree level, weak force mediated standard model decay process: muon decay, $\mu^-\rightarrow \nu_\mu + \bar{\nu}_e + e^-$, as the process are very similar (see Figure~\ref{fig:diagrams}).  

For muons with a Lorentz factor $\gamma_{\mu}$ in the observer's frame the decay rate is found to be
\begin{equation}
{ \Gamma \  =   \gamma_{\mu}^{-1}} {{G_F^2 m_{\mu}^5}\over{192\pi^3}}
\label{mu}
\end{equation}
where $G_F^2 = g^4/32M_{W}^4$, is the square of the Fermi constant equal to $1.360 \times 10^{-10} \ {\rm GeV}^{-4}$, with $g$ being the weak coupling constant and $M_{W}$ being the $W$-boson mass in electroweak theory.

We can now apply our effective energy-dependent mass-squared formalism given by equation (\ref{effectivemass}) to determine the scaling of the emission rate with the $\delta$ parameter and with energy. Noting that for any reasonable neutrino mass, $m_{\nu} \ll 2\delta_{\nu e} E_{\nu}^2$, it follows that 
$\tilde{m}^2(E) \simeq 2\delta_{\nu e} E_{\nu}^2$. 
We therefore make the substitution
\begin{equation}
m_{\mu}^2 \ \rightarrow \ 2\delta_{\nu e} E_{\nu}^2 
\end{equation}
\noindent from which it follows that
\begin{equation}
{\gamma_{\mu}^2} \ \rightarrow \ {{E_{\nu}^2}\over{2\delta_{\nu e} E_{\nu}^2}} = (2\delta_{\nu e})^{-1}.
\end{equation}
The rate for the vacuum pair emission processes (VPE) is then
\begin{equation}
\Gamma \ \propto \ (2\delta_{\nu e})^{1/2} G_F^2 (2\delta_{\nu e} E_{\nu}^2)^{5/2}  
\label{dimension}
\end{equation}
\noindent which gives the proportionality
\begin{equation}
\Gamma \ \propto \ G_F^2 \ \delta_{\nu e}^3 E_{\nu}^5 
\label{prop}
\end{equation}
\noindent showing the strong dependence of the decay rate on both $\delta_{\nu e}$ and $E_{\nu}$.  The upper limit on $\delta_{e}$ in the $[d] = 4$ case was  obtained in Ref.~\cite{st14b}.
These dependences are in agreement with those given in Refs.~\cite{co11} and~\cite{ma13} and the only unknown is the numerical coefficient, the calculation of which we now address.
\subsection{Decay by vacuum electron-positron pair emission}

Above an energy threshold given by
\begin{equation}
E_{th} = m_e\sqrt{{{2}\over{{\delta}}}} 
\label{eq:threshold}
\end{equation}
\cite{sg01} with $\delta \equiv \delta_{\nu e}$ given by equation (\ref{sub}) the rate for the VPE process, $\nu \to \nu \,e^+\, e^-$ via the neutral current (NC) $Z$-exchange channel, has been calculated by~\cite{co11} to be
\begin{equation}
\Gamma = \frac{1}{14}\frac{G_F^2 (2\delta)^3E_{\nu}^5}{192\,\pi^3} = 1.31 \times 10^{-14} \delta^3 E_{\rm GeV}^5\ \ {\rm GeV}.
\label{G}
\end{equation}

In general, the charged current (CC) $W$-exchange channels contribute as well.  However, this channel is only kinematically relevant for $\nu_e$'s, as the production of $\mu$ or $\tau$ leptons by $\nu_{\mu}$'s or $\nu_{\tau}$'s has a much higher energy threshold due to the large final state particle masses (equation (\ref{eq:threshold}) with $m_e$ replaced by $m_{\mu}$ or 
$m_{\tau}$) and our final results are highly threshold dependent. Owing to neutrino oscillations, neutrinos propagating over large distances spend 1/3 of their time in each flavor state. Thus, the flavor population of neutrinos from astrophysical sources is expected to be [$\nu_e$:$\nu_{\mu}$:$\nu_{\tau}$] = [1:1:1] so that CC interactions involving $\nu_e$'s will only be important 1/3 of the time. We ignore CC interactions involving $\nu_e$'s in our calculations but discuss their impact on our conclusions later.

The mean fractional energy loss due to a VPE is $\sim$ 0.78~\cite{co11}.\footnote{The vacuum \v{C}erenkov emission (VCE) process, $\nu \rightarrow \nu + \gamma$, is also kinematically allowed for superluminal neutrinos. However, since the neutrino has no charge, this process entails the neutral current channel production of a virtual electron-positron pair followed by its annihilation into a photon. Thus, the rate for VCE is a factor of $\alpha$ lower than that for VPE.} 
Using equations (\ref{sub}) and (\ref{prop}) and the dynamical matrix element taken from the simplest case (example 1) derived in Ref.~\cite{ca12}, we can generalize equation (\ref{G}) to $n = 1$ and $n = 2$
\begin{equation}
\Gamma = \frac{G_F^2}{192\,\pi^3}[(1-2s_W^2)^2 + (2s_W^2)^2]\zeta_n \kappa_n^3 \frac{E_\nu^{3n+5}}{M_{Pl}^{3n}} 
\label{G2}
\end{equation} 
\noindent where $s_W$ is the sine of the Weinberg angle ($s_W^2 = 0.231$) and the $\zeta_n$'s are  numbers of order $1$~\cite{ca12}. 

For the $n = 1$ case we obtain the VPE rate
\begin{equation}
\Gamma = 1.72 \times 10^{-14} \kappa_1^3 E_{\rm GeV}^5\ (E/M_{Pl})^3 \ {\rm GeV},
\label{Gn1}
\end{equation}
and for the $n = 2$ case we obtain the VPE rate
\begin{equation}
\Gamma = 1.91 \times 10^{-14} \kappa_2^3 E_{\rm GeV}^5\ (E/M_{Pl})^6 \ {\rm GeV}.
\label{Gn2}
\end{equation}
 
\subsection{Decay by neutrino pair emission (neutrino splitting)}

The process of neutrino
splitting in the case of superluminal neutrinos, i.e., $\nu \rightarrow 3\nu$ is relatively unimportant in the $[d] = 4, n = 0$ case
considered in Ref.~\cite{st14} because neutrinos of comparable energy but different flavor 
travel at virtually the same velocity, as indicated by neutrino oscillation experiments~\cite{co11,ab14}.
In the presence of $[d] > 4$ $(n > 0)$ terms in a Planck-mass suppressed EFT, the
energy dependent velocity differences in the $n > 0$ cases become significant~\cite{ma13}. Superluminal neutrino splitting becomes kinematically allowed because of the dependence of velocity on energy. The daughter neutrinos travel with a smaller velocity. The velocity dependent energy of the parent neutrino is therefore greater than that of the daughter neutrinos.  We therefore consider the $n = 1$ and $n = 2$ scenarios in this paper with particular regard to using the IceCube neutrino observations~\cite{aa14} to place constraints on superluminality in the neutrino sector.

The neutrino splitting is an NC interaction. The total neutrino splitting
rate obtained is therefore three times that of the NC mediated VPE process above threshold. We assume that the three daughter neutrinos
each carry off approximately 1/3 of the energy of the incoming neutrino. Therefore,
for the $n = 1$ case we obtain the neutrino splitting rate
\begin{equation}
\Gamma = 5.16 \times 10^{-14} \kappa_1^3 E_{\rm GeV}^5\ (E/M_{Pl})^3 \ {\rm GeV},
\label{Gsplitn1}
\end{equation}
and for the $n = 2$ case we obtain the neutrino splitting rate
\begin{equation}
\Gamma = 5.73 \times 10^{-14} \kappa_2^3 E_{\rm GeV}^5\ (E/M_{Pl})^6 \ {\rm GeV}.
\label{Gsplitn2}
\end{equation}

The threshold energy for neutrino splitting is proportional to the neutrino mass and is, in any case, much smaller than 100 TeV. We can therefore assume that we are always above threshold when comparing with the IceCube data.

\section{The neutrinos observed by IceCube} 

As of this writing, the IceCube collaboration has identified $87_{-10}^{+14}$ events from neutrinos
of astrophysical origin with energies above 10 TeV, with the error in the number of astrophysical events determined by the modeled subtraction of both conventional and prompt atmospheric neutrinos and also penetrating atmospheric muons, particularly at energies below 60 TeV~\cite{aa14b}. Neutrinos identified to be of astrophysical origin and having energies above 60 TeV were found to have an energy spectrum proportional to $E_{\nu}^{-2}$~\cite{aa14,aa14b,aa13}. 

There are are four indications that the the bulk of cosmic neutrinos observed by IceCube with energies above 0.1 PeV are extragalactic in origin: (1) The arrival distribution of the 37 reported events with $E > 0.1$ PeV observed by IceCube above atmospheric background is consistent with isotropy, with no significant enhancement in the galactic plane~\cite{aa14},
(2) At least one of the ~PeV neutrinos came from a direction off the galactic plane~\cite{aa14},
(3) The diffuse galactic neutrino flux~\cite{st79} is expected to be well below that observed by  IceCube, (4) Upper limits on diffuse galactic $\gamma$-rays in the TeV-PeV energy range imply that galactic neutrinos cannot account for the neutrino flux observed by IceCube~\cite{ah14}.

Above 60 TeV, the IceCube data are consistent with a spectrum given by $E_{\nu}^2(dN_{\nu}/dE_{\nu}) \simeq \ 10^{-8} \ {\rm GeV}{\rm cm}^{-2}{\rm s}^{-1}$. Spectra steeper than $E_{\nu}^{-2}$ do not give a good fit to the existing data in the 60 TeV to 2 PeV energy range~\cite{ch14}. However, no neutrino induced events have been seen above $ \sim 2$ PeV, as would be expected from extending an $E_{\nu}^{-2}$ spectrum beyond $\sim 2$ PeV. In particular, IceCube has not detected any neutrino induced events from the Glashow resonance effect at 6.3 PeV. In this effect, electrons in the IceCube volume provide enhanced target cross sections for $\bar{\nu}_{e}$'s through the $W^-$ resonance channel, $\bar{\nu}_{e} + e^- \rightarrow W^- \rightarrow shower$, at the resonance energy $E_{\bar{\nu}_{e}} = M_W^2/2m_{e} = 6.3$ PeV~\cite{gl60}. This enhancement leads to an increased IceCube effective area for detecting the sum of the ${\nu}_{e}$'s, {\it i.e.}, ${\nu}_{e}$'s plus $\bar{\nu}_{e}$'s by a factor of $\sim 10$~\cite{aa13}. It is usually expected that 1/3 of the potential 6.3 PeV neutrinos would be ${\nu}_{e}$'s plus $\bar{\nu}_{e}$'s unless new physics is involved.
Thus, the enhancement in the overall effective area expected is a factor of $\sim$3. Taking account of the increased effective area between 2 and 6 PeV and a decrease from an assumed neutrino energy spectrum of $E_{\nu}^{-2}$, we would expect about 3 events at the Glashow resonance provided that the number of $\bar{\nu}_{e}$'s is equal to the number of ${\nu}_{e}$'s. Even without considering the Glashow resonance effect, several neutrino events above 2 PeV would be expected if the $E_{\nu}^{-2}$ spectrum extended to higher energies. Thus, the lack of neutrinos above 2 PeV energy and at the 6.3 PeV resonance may be indications of a cutoff in the neutrino spectrum. 

\section{Calculations of Superluminal Neutrino Propagation with $[d] > 4$ Operator Dominance}

We have used Monte Carlo techniques to determine the effect of neutrino splitting and VPE on putative superluminal neutrinos propagating from cosmological distances under the assumption of the dominance of Planck mass suppressed LIV operators with $[d] > 4$. Our Monte Carlo codes take account of energy losses by both neutrino splitting and VPE as well as redshifting of neutrinos emitted from sources at cosmological distances. As in Ref.~\cite{st14}, we consider a scenario where the neutrino sources have a redshift distribution that follows that of the star formation rate~\cite{be13}. This redshift distribution appears to be roughly applicable for both active galactic nuclei and $\gamma$-ray bursts. We assume a simple source spectrum proportional to $E^{-2}$ between 100 TeV and 100 PeV
as is the case for cosmic neutrinos observed by IceCube with energies above 60 TeV~\cite{aa14}. We generate Monte Carlo events using these two distributions.  Our final results are normalized to an energy flux of $E_{\nu}^2(dN_{\nu}/dE_{\nu}) \simeq 10^{-8}\  {\rm GeV}{\rm cm}^{-2}{\rm s}^{-1}{\rm sr}^{-1}$, as is consistent with the IceCube data for both the southern and northern hemisphere for energies between 60 TeV and 2 PeV.~\cite{aa14}. In our Monte Carlo runs we considered VPE threshold energies between 10 PeV and 40 PeV for the VPE process, corresponding to values of $\delta_{\nu e}$ between $5.2\times10^{-21} \ {\rm and}~ 3.3 \times 10^{-22}$. By propagating our test neutrinos including energy losses from VPE, neutrino splitting, and redshifting using our Monte Carlo code, we obtained final neutrino spectra and compared them with the IceCube results.

Given that neutrinos detected by IceCube are extragalactic, cosmological effects should be taken into account in deriving new LIV constraints. The reasons are straightforward. As opposed to the extinction of high energy extragalactic photons through electromagnetic interactions~\cite{st92}, neutrinos survive from all redshifts because they only interact weakly.  It follows that since the universe is transparent to neutrinos, most of the cosmic PeV neutrinos will come from sources at redshifts between $\sim$0.5 and $\sim$2~\cite{be13}. Therefore, along with energy losses by VPE~\cite{co11} and neutrino splitting, energy losses by redshifting of neutrinos and the effect of the cosmological $\Lambda$CDM redshift-distance relation 
\beq
D = {{c}\over{H_0}}\int\limits_0^z\frac{dz'}{(1+z')\sqrt{\Omega_{\rm\Lambda} + \Omega_{\rm M} (1 + z')^3}}
\eeq 
need to be included in the determination of $\delta_{\nu}$. 

As in Ref.~\cite{st14}, we assume a flat $\Lambda$CDM universe with a Hubble constant of $H_0 =$ 67.8 km s$^{-1}$ Mpc$^{-1}$ along with $\Omega_{\rm\Lambda}$ = 0.7 and $\Omega_{\rm M}$ = 0.3.  Thus the energy loss due to redshifting is given by
\begin{equation}
-(\partial \log  E/\partial t)_{redshift} =  H_{0}\sqrt{\Omega_{m}(1+z)^3 +
  \Omega_{\Lambda}}.
\label{redshift}  
\end{equation}
The decay widths for the VPE process are given by equations (\ref{Gn1}) and (\ref{Gn2}) for the cases $n = 1$ and $n = 2$ respectively while those for neutrino splitting are given by equations (\ref{Gsplitn1}) and (\ref{Gsplitn2}). 

\section{RESULTS}

\subsection{[d] = 6 $\cal{CPT}$ Conserving Operator Dominance} 

\begin{figure}[!t]
{\includegraphics[scale=0.6]{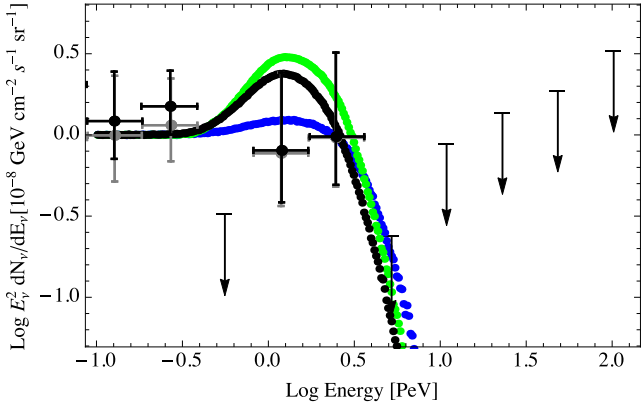}}
\caption{Separately calculated n = 2 neutrino spectra with the VPE case shown in blue and the neutrino splitting case shown in green. The black spectrum takes account of all three processes
(redshifting, neutrino splitting, and VPE) occurring simultaneously. The rates for all cases are fixed by setting the rest frame threshold energy for VPE at 10 PeV. The neutrino spectra are normalized to the IceCube data both with (gray) and without (black) an estimated flux of prompt atmospheric neutrinos subtracted.~\cite{aa14}.}
\label{combined}
\end{figure}

As found before~\cite{st14}, the best fit to the IceCube data corresponds to a VPE rest-frame threshold energy $E_{\nu, \rm th} = 10$ PeV as shown in Figure~\ref{combined}. This corresponds to
$\delta_{\nu e} \equiv \delta_{\nu} - \delta_e \le \ 5.2 \times 10^{-21}$. Noting that $\delta_{e} \le 5 \times 10^{-21}$~\cite{st14b}, we found previously that $\delta_{\nu} \le 1.0 \times 10^{-20}$. Should we assume that $\delta_{e}$ is negligible compared to $\delta_{\nu}$~\cite{st14b} then $\delta_{\nu} \simeq \delta_{\nu e}$. We note that one can not assume that $\delta_{\nu}$ and $\delta_e$ are equal. Models can be constructed where $\delta_{\nu}$ and $\delta_e$ are independent and it has even been suggested that LIV may occur only in the neutrino sector~\cite{ca12}.   

Values of $E_{\nu, \rm th}$ less than 10 PeV are inconsistent with the IceCube data. The result for a 10 PeV rest-frame threshold energy, corresponding to $\delta_{\nu e} = \ 5.2 \times 10^{-21}$, is just consistent with the IceCube results, giving a cutoff effect above 2 PeV. Thus for the conservative case of no-LIV effect, {\it e.g.}, if one assumes a cutoff in the intrinsic neutrino spectrum of the sources, or one assumes a slightly steeper PeV-range neutrino spectrum proportional to $E_{\nu}^{-2.3}$, we previously obtained the constraint on superluminal neutrino velocity, $\delta_{\nu} = \delta_{\nu e} + \delta_e  \le \ 1.0 \times 10^{-20}$~\cite{st14}.

In the case of the $\cal{CPT}$ conserving [d] = 6 operator (n = 2) dominance, the results in Figure~\ref{combined} show a high-energy drop off in the propagated neutrino spectrum near the redshifted VPE threshold energy and a pileup in the spectrum below that energy. This predicted drop off may be a possible explanation for the lack of observed neutrinos above 2 PeV (see Section V) as suggested previously~\cite{st14}. This pileup is caused by the propagation of the higher energy neutrinos in energy space down to energies within a factor of $\sim$5 below the VPE threshold. This is indicative of the fact that fractional energy loss from the last allowed neutrino decay before the VPE process ceases is 0.78~\cite{co11} and that for neutrino splitting is taken to be 1/3. The pileup effect is similar to that of energy propagation for ultrahigh energy protons near the GZK threshold~\cite{st89}.

The pileup effect caused by the neutrino splitting process is more pronounced than that caused by the VPE process because neutrino splitting produces two new lower energy neutrinos per interaction. This would be a way of distinguishing a dominance of $[d] > 4$ Planck-mass suppressed interactions from $[d] = 4$ interactions. Thus, with better statistics in the energy range above 100 TeV,
a significant pileup effect would be a signal of Planck-scale physics.

\begin{figure}[!t]
{\includegraphics[scale=0.6]{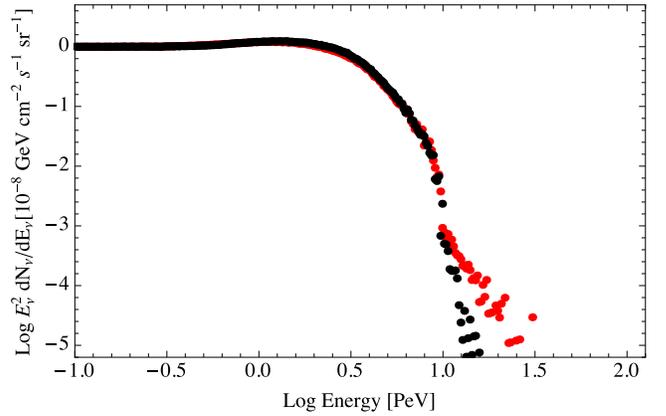}}
\caption{Calculated n = 0  (red) and n = 2 (black, as in Figure \ref{combined}) neutrino spectrum obtained for the VPE process only (no neutrino splitting) simultaneously with redshifting. The rates for all cases are fixed by setting the threshold energy for VPE at 10 PeV}
\label{vpeonly}
\end{figure}

\begin{figure}[!t]
{\includegraphics[scale=0.6]{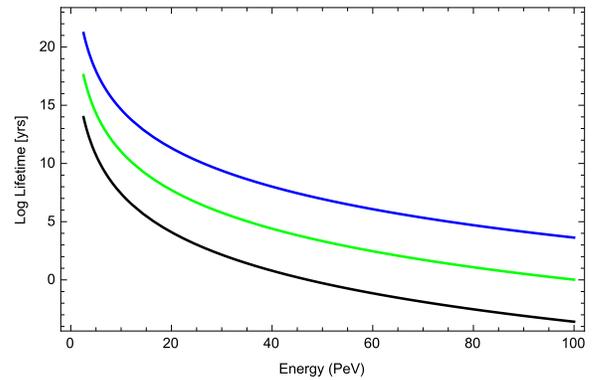}}
\caption{Mean decay times for neutrino splitting process in the n = 2 case obtained by setting the threshold energy for VPE at 10 PeV (black), 20 PeV (green), and 40 PeV (blue).}
\label{lifetimes}
\end{figure}

\begin{figure}[!t]
{\includegraphics[scale=0.6]{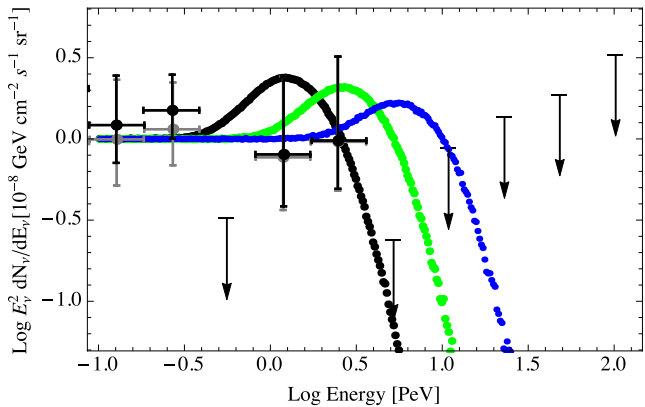}}
\caption{Calculated n = 2 spectra taking into account of all three processes
(redshifting, neutrino splitting, and VPE) occurring simultaneously for rest frame VPE threshold energies  of 10 PeV (black, as in Figure \ref{combined}), 20 PeV (green), and 40 PeV (blue). The IceCube data
are as in Figure~\ref{combined}~\cite{aa14}.}
\label{thresholdeffects}
\end{figure}

In order to test for threshold effects in the VPE process, we employed a Monte Carlo routine to find the opening up of phase space.  We assume the same LIV parameters for every particle but with an electron mass for two of the outgoing states.  We find that the entirety of phase space is available when the energy reaches about 1.6 times that of threshold.  Threshold effects should therefore have little impact on our results, as above this value full rates are operative.  In practice, neutrinos near threshold rarely pair produce before dropping below this energy due to redshifting since their mean decay times increase with their decreasing energy as they propagate.  A Monte Carlo exploration of phase space for neutrino splitting yields similar results however the threshold for this reaction is in the GeV range meaning that full rates apply throughout our calculation. This also justifies our assumption that the neutrino splitting and VPE rates are similar per decay channel.

Throughout our calculation we have assumed that a neutrino loses 0.78 of its initial energy per VPE interaction.  Equation (\ref{G2}) shows that the VPE rates 
do not differ by more than 45\% between the $n = 0$ and 
$n = 2$ cases. This reflects the difference in the phase space 
factors, since the dynamical matrix elements are the same, indicating that this is also the 
maximum deviation in the fraction of energy carried off by the neutrino 
in VPE. It is likely that the deviation would be at most a third of that in a
three-body decay, {\it viz.}, 15\% meaning that the resulting energy fraction for the 
$n = 2$ case could be as high as 0.25.  We tested this and found that it produces no discernible difference in the spectra.  We also tested an energy fraction of 0.5 and found that even this extreme case would generate no observational consequences on the pileup effect. 

In Figure~\ref{vpeonly}, we plot the VPE process alone (along with redshifting) for the $\cal{CPT}$- conserving cases $n = 0$ and $n = 2$.  We see that the resulting spectra are indistinguishable below threshold.  Events above the redshifted threshold pair produce in relatively short times compared to cosmological timescales regardless of the energy dependence, making the spectra for $n = 0$ and $n = 2$ below the redshifted threshold indistinguishable.  We can only see the expected differences in the steepening of the spectra for energies above threshold owing to the rate differences between
$n = 0$ and $n = 2$ given by equation (\ref{G2}).

As can be seen in Figure ~\ref{lifetimes}, the mean decay times increase for the neutrino splitting process with increasing choice of VPE threshold.  The increased mean decay times have the effect of reducing the pileup for increased choice of threshold as fewer neutrino splitting events will occur.  Thus the pileup becomes a somewhat less sensitive test of Planck-scale effects with increasing threshold energies. Figure~\ref{thresholdeffects} shows the effects of choosing different threshold energies. The dominant process continues to be that of neutrino splitting but with decreasing importance.

\subsection{[d] = 5 \cal{CPT} Violating Operator Dominance}

In the n = 1 case, the dominant $[d] = 5$ operator violates $\cal{CPT}$.
Thus, if the $\nu$ is superluminal, the $\bar{\nu}$ will be subluminal, and {\it vice
versa}. However, the IceCube detector cannot distinguish neutrinos from antineutrinos. 
The incoming $\nu (\bar{\nu}$) generates a shower in the detector, allowing
a measurement of its energy and direction. Even in cases where there is a muon
track, the charge of the muon is not determined. 

There would be an exception for electron antineutrinos at 6.3 PeV, 
given an expected enhancement in the event rate at the $W^{-}$ Glashow 
resonance since this resonance only occurs with $\bar{\nu_{e}}$. 
However, as we have discussed, no events have been detected
above 2 PeV. We note that $\nu - \bar{\nu}$ oscillation measurements would give
the strongest constraints on the difference in $\delta$'s between $\nu$'s and 
$\bar{\nu}$'s~\cite{ab14}. 

Since both VPE and neutrino splitting interactions generate
a particle-antiparticle lepton pair, one of the pair particles will be
superluminal ($\delta > 0$) whereas the other particle will be subluminal 
($\delta < 0$)~\cite{km13}. Thus, of the daughter particles,
one will be superluminal and interact, while the other will only redshift.
We have accounted for this in our simulations. 

\begin{figure}[!t]
{\includegraphics[scale=0.6]{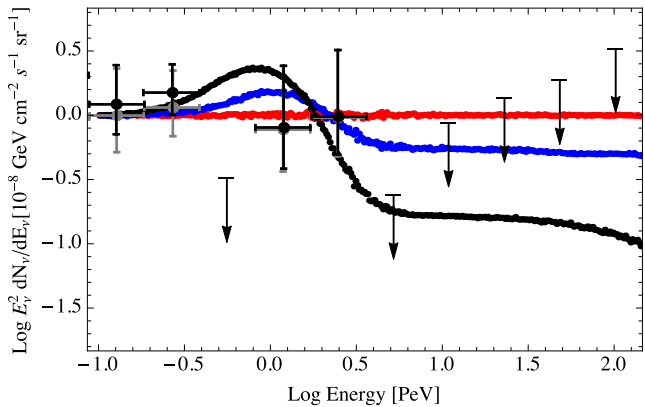}}
\caption{Calculated n = 1 neutrino spectra assuming 100\% (black), 50\% (blue) and 0\% (red) initial superluminal neutrinos (antineutrinos). The neutrino spectra are normalized to the IceCube
data~\cite{aa14}.}
\label{CPTodd}
\end{figure}

Figure~\ref{CPTodd} shows the results in the $\cal{CPT}$-violating $n = 1$ case, assuming 100\%, 50\% and 0\% initial superluminal neutrinos (antineutrinos) and propagating the spectrum using our Monte Carlo program and taking account of the fact that in all cases, one of the daughter leptons is subluminal and therefore does not undergo further interactions. As a sanity check, we see that in the 0\% case only redshifting occurs, preserving the initial $E^{-2}$ spectrum. The other cases show the effect of VPE and neutrino splitting by both the initial fraction of superluminal neutrinos and the superluminal daughter neutrinos.  

Thus, as opposed to the
$\cal{CPT}$-conserving $n = 2$ case,  no clearly observable cut off is produced, with the possible
unrealistic exception of postulating that only superluminal $\nu$'s (or superluminal $\bar{\nu}$'s)
are produced in cosmic sources. That case, shown in black in Fig.~\ref{CPTodd}, as well as the
other case of postulating no initial superluminal neutrinos, shown in red, are shown for illustrative purposes. The 50/50 case, shown in blue, is more realistic. 

We note that in the n = 1, $\cal{CPT}$-odd case, the
details of the kinematics are different from the n = 0 and n = 2, $\cal{CPT}$-even
cases because in the $\cal{CPT}$-odd case the signs of $\delta$ are
opposite for $\nu$'s and $\bar{\nu}$'s. If we assume that they are equal and opposite, then the rate given in equation (\ref{G2}) would maximally be altered by replacing the $\delta$ with 2$\delta$.  Since the source kinematics dominate as the daughter energies are comparable, doubling delta should overestimate their contribution to the overall rate. We have applied Monte Carlo techniques to explore the phase space and find that the subliminal particle will carry away a slightly higher fraction of the energy after the split ($\sim 40$\%) in the $\cal{CPT}$-odd case. By making these modifications to our code we find that there is little observational difference between the modified results and those obtained assuming the same rate as given by equation (\ref{G2}).  An exact treatment of the kinematics for $\cal{CPT}$-odd, which are complex, are therefore unnecessary and our spectral results in the $\cal{CPT}$-odd case given in Figure \ref{CPTodd} are a good approximation to an exact treatment.

\section{CONCLUSIONS}

In this paper, we have explored the effects of $[d] > 4$ Planck-mass suppressed operators on the propagation and resulting energy spectrum of superluminal neutrinos of extragalactic origin. We have expressed these Lorentz violating perturbations as a modifications of the energy-momentum dispersion relation in the form  $\delta_{\nu e} \simeq \delta_{\nu} \equiv \delta_n$ (see discussion in IIIA) for Planck mass suppressed energy dependent values of $\delta_{n}$ as defined in equation (\ref{sub}). These terms can arise from higher dimension operators in the EFT formalism~\cite{ck98}. 

We find that a high-energy drop off in a propagated superluminal neutrino spectrum above $\sim 2$ PeV results from kinematically allowed weak neutral current processes in the $\cal{CPT}$-conserving cases. The drop off matches the observed neutrino spectra for energy dependent values of $\delta_n$ fixed to be $5.2 \times 10^{-21}$ at 10 PeV as shown in Figure \ref{combined}. This implies a required value for $\kappa_2$ of $7.8 \times 10^3$ with $\kappa_{2}$ as defined in equation (\ref{sub}). Our new results apply directly to both the $[d] = 4$ case as found before~\cite{st14} and to the $[d] = 6$ operator in the SME~\cite{km13,di14}. These values are very well defined by the fit shown in Figure \ref{combined} combined with the strong functional relation $\delta \propto E_{th}^{-2}$.

We have not included the effect of CC interactions involving $\nu_e$'s in our calculations. However, as noted in Section IIIA, the contribution from CC interactions is flavor suppressed by a factor of three relative to the NC channel.  There is again further flavor suppression of VPE relative to neutrino splitting, as neutrino splitting involves three possible final state neutrino flavor decay channels with negligible velocity differences~\cite{ab14} whereas VPE involves only the electronic sector~\cite{ma13}. Hence neutrino splitting becomes the dominant energy loss mechanism and the charged current contribution to the VPE rate is a small correction to the overall observational signal. This correction will not affect the cutoff energy, but will only produce a
small, presently unobservable, contribution to the pileup effect.

We further note that, should the $\cal{CPT}$-violating [d] = 5 operator dominate, we would find an absence of a clear cutoff in the propagated neutrino spectrum. Since this result contradicts our thesis that the observed 2 PeV drop off in the neutrino spectrum may be due to Planck-scale physics,   we can conclude that, within this framework, the dominant term in the EFT can not be a $\cal{CPT}$-violating the dimension-5 operator. 

In the SME EFT formalism, our results have significant quantitative implications: If the cutoff above $\sim 2$ PeV in the neutrino spectrum is caused by LIV, this would result from an EFT with either a dominant dimension-4 term with $\mathaccent'27 c^{(4)} = -\delta_{\nu e} = 5.2 \times 10^{-21}$ as given in Ref.~\cite{st14b}, or by a dominant SME dimension-6 term with $\mathaccent'27 c^{(6)} = -\kappa_2/M_{Pl}^2 \ge - 5.2 \times 10^{-35}$ GeV$^{-2}$. We further find that the pileup feature is more pronounced in the case of $[d] = 6$ operator dominance than in the $[d] = 4$ case. The detection of a pronounced pileup feature just below a $\sim 2$ PeV cutoff energy would require the detection of many more astrophysical neutrinos above 100 TeV. However, the detection of such a pronounced pileup together with a cutoff would be {\it prima facie} evidence of $\cal{CPT}$-even LIV that becomes strong $\sim 2$ orders of magnitude below the Planck scale. In this regard, we note that $\cal{CPT}$-even LIV in the gravitational sector at energies below the Planck energy has been considered in the context of Ho\v{r}ava-Lifshitz gravity~\cite{po12}, thus allowing a potential theoretical basis for our EFT analysis of the IceCube observations.  

On the other hand, if the cutoff is caused by a natural break in the neutrino spectra of the astrophysical neutrino sources, the above numbers for the $\mathaccent'27 c^{(4)}$ and $\mathaccent'27 c^{(6)}$ SME coefficients then become the best limits on these values. Such limits would be significantly better than those derived in Ref.~\cite{di14} because we realistically take account of the redshift distribution of extragalactic neutrino sources and we therefore find a higher effective rest-frame threshold energy.

\section*{Acknowledgments} 
We thank Andrew Cohen and Alan Kosteleck\'{y} for helpful comments.
S. Liberati acknowledges support of a grant from the John Templeton Foundation.


\begin{thebibliography}{99}

\bibitem{pl99} M. Planck, Mitt. Thermodyn., Folg. {\bf 5} (1899).

\bibitem{do95} S. Doplicher, K. Fredenhagen and J. E. Roberts, Commun. Math. Phys. {\bf 172}, 187 (1995).

\bibitem{ko89} V. A. Kosteleck\'{y} and S. Samuel, Phys. Rev. D {\bf 39}, 683 (1989).

\bibitem{dm05} D.\ Mattingly, Liv.\ Rev.\ Rel.\ {\bf 8}, 5 (2005).

\bibitem{li13} S. Liberati, Class. Quantum Grav. {\bf 30}, 133001 (2013).

\bibitem{aa14} M. G. Aartsen {\it et al.} (IceCube), Phys. Rev. Lett. {\bf 113}, 101101 (2014).

\bibitem{co99} S. R. Coleman and S. L. Glashow, Phys. Rev. D {\bf 59}, 
116008 (1999).

\bibitem{ck98} D. Calladay and V. A. Kosteleck\'{y}, Phys. Rev. D {\bf 58}, 116002 (1998).

\bibitem{ho09} P. Ho\v{r}ava, Phys. Rev. D {\bf 79}, 084008 (2009). 

\bibitem{ko12} V. A. Kosteleck\'{y} and M. Mewes, Phys. Rev. D {\bf 85}, 096005 (2012).
 
\bibitem{st14} F. W. Stecker and S. T. Scully, Phys. Rev. D {\bf 90}, 043012 (2014).

\bibitem{gr02} O. W. Greenberg, Phys. Rev. Lett. {\bf 89}, 231602 (2002).

\bibitem{ko09}  V. A. Kosteleck\'{y} and M. Mewes, Phys. Rev. D {\bf 80}, 015020 (2009).

\bibitem{ja03} T. Jacobson, S. Liberati, and D. Mattingly, Nature {\bf 424}, 1019 (2003);
T. Jacobson, S. Liberati, and D. Mattingly, and F. W. Stecker, Phys. Rev. Letters {\bf 93}, 021101 (2004); R. Montemayor and L. F. Urruta,  Phys. Rev. D {\bf 72}, 045018 (2005); B. Altschul, Phys. Rev. D {\bf 74}, 083003 (2006); L. Maccione, S. Liberati, A. Celotti and J. Kirk, JCAP {\bf 0710}, 013 (2007). 

\bibitem{st14b} F. W. Stecker, Astropart. Phys. {\bf 56}, 16 (2014).

\bibitem{ab14} K. Abe, {\it et al.}, arXiv:1410.4267 (2014).

\bibitem{co11} A. G. Cohen and S. L. Glashow, Phys. Rev. Lett. {\bf 107}, 181803 (2011).

\bibitem{ma13} L. Maccione, S. Liberati and D. Mattingly, JCAP {\bf 03}, 039 (2013).

\bibitem{sg01} F.W. Stecker and S.L. Glashow, Astropart. Phys. {\bf 16}, 97 (2001).

\bibitem{ca12} J. M. Carmona, J. L. Cort\'{e}s and D. Maz\'{o}n, Phys. Rev. D {\bf 85}, 113001 (2012).

\bibitem{aa14b} M. G. Aartsen {\it et al.} (IceCube), Phys. Rev. D {\bf 91}, 022001 (2015).

\bibitem{aa13} M. G. Aartsen {\it et al.} (IceCube), Science {\bf 342}, 1242856 (2013).

\bibitem{st79} F. W. Stecker, Astrophys. J. {\bf 228}, 919 (1979).

\bibitem{ah14} M. Ahlers and K. Murase, Phys. Rev. D {\bf 90}, 023010 (2014).

\bibitem{ch14} C.-Y. Chen, P. S. Bhupal Dev and A. Soni,  Phys. Rev. D {\bf 89}, 033012 (2014).

\bibitem{gl60} S. L. Glashow, Phys. Rev. {\bf 118}, 316 (1960).

\bibitem{be13} P. S. Behroozi, R. H. Wechsler and C. Conroy, Astrophys. J. {\bf 770}:57 (2013).

\bibitem{st92} F. W. Stecker,O. C. de Jager and M. H. Salamon, Astrophys. J. Letters {\bf 390}, L49
(1992).

\bibitem{st89} F.W. Stecker, {\it Nature} {\bf 342}, 401 (1989).

\bibitem{km13} V. A. Kosteleck\'{y} and M. Mewes, Phys. Rev. D {\bf 88}, 096006 (2013).

\bibitem{di14} J. S. Diaz, V. A. Kosteleck\'{y} and M. Mewes, Phys. Rev. D {\bf 89}, 043005 (2014).

\bibitem{po12} M. Pospelov and Y. Shang, Phys. Rev. D {\bf 85}, 105001 (2012).






 
\end{thebibliography}
\end{document}